\documentclass[journal,twoside,web]{ieeecolor}
\usepackage{generic}
\def\logoname{blank_logo}
\usepackage{cite}
\usepackage{amsmath,amssymb,amsfonts}
\usepackage{graphicx}
\usepackage{epstopdf}
\usepackage{hyperref}
\hypersetup{hidelinks=true}
\usepackage[table]{xcolor}
\usepackage{orcidlink}

\def\red#1{\protect\textcolor{red}{#1}}

\usepackage{soul}
\newcommand{\grnhl}[1]{\begingroup\sethlcolor{green!40}\hl{#1}\endgroup}
\def\rsn#1{\protect\textcolor{black}{\protect\grnhl{#1}}}
\sethlcolor{yellow}
\soulregister\cite{7}
\soulregister\eqref{7}
\newcommand{\yel}[1]{\hl{#1}}
\soulregister\red{7}
\soulregister\yel{7}
\soulregister\rsn{7}
\usepackage{textcomp}
\usepackage{booktabs}
\usepackage{multirow}
\usepackage{threeparttable}
\usepackage{array}
\usepackage[caption=false,font=normalsize,labelfont=sf,textfont=sf]{subfig}
\usepackage{etoolbox}
\def\BibTeX{{\rm B\kern-.05em{\sc i\kern-.025em b}\kern-.08em
    T\kern-.1667em\lower.7ex\hbox{E}\kern-.125emX}}

\makeatletter
\newlength{\fullheadwidth}
\def\ps@titlepagestyle{\def\@oddfoot{}\def\@evenfoot{}%
  \def\@oddhead{\vbox{\hbox to\fullheadwidth{\scriptsize\sffamily\leftmark\hfil\thepage}%
    \vspace*{1.5pt}\hrule width\fullheadwidth height1pt}}%
  \def\@evenhead{\vbox{\hbox to\fullheadwidth{\scriptsize\sffamily\thepage\hfil\leftmark}%
    \vspace*{1.5pt}\hrule width\fullheadwidth height1pt}}}
\makeatother

\selectfont
\begin{document}

\title{An intact blood-wall interface as a transport readout of ejection fraction in echocardiography}

\author{Xuan~Lin\,\orcidlink{0009-0000-3081-8442}, Lu Chen\orcidlink{0000-0003-0351-7980}, and Binjie~Qin\orcidlink{0000-0001-7445-1582},~\IEEEmembership{Member,~IEEE}
\thanks{The manuscript was received on June 10, 2026, and accepted on October 6, 2024. This work was partially supported by the Science and Technology Commission of Shanghai Municipality (19dz1200500, 19411951507) and the Interdisciplinary Program of Shanghai Jiao Tong University (ZH2018ZDA19, YG2021QN122, YG2021QN99). X. Lin, and B. Qin are with the School of Biomedical Engineering, Shanghai Jiao Tong University, Shanghai 200240, China. Lu Chen is with the Center for Cardiovascular Medicine, Shanghai Chest Hospital, Shanghai Jiao Tong University School of Medicine, Shanghai, China. She is also with Department of Cardiology, The First Affiliated Hospital of Soochow University, Suzhou, China. (Corresponding author: Binjie Qin, Email: bjqin@sjtu.edu.cn)}}

\maketitle

\begin{abstract}
Ejection fraction (EF) compresses the blood-wall interface's per-cycle kinematics into one scalar. Two conventions derive it from static contours or regress it from pixels. Both cut the intact interface carrying that motion, discard the phase trajectory and coupling, and pay for the loss in labelled cohorts and compute. We instead read EF out as a compressed transport across three levels: an outer level, through a bidirectional gated recurrent unit (GRU) and a continuous ordinary-differential-equation (ODE) block, carries high-dimensional motion onto a one-dimensional readout in closed loop with end-diastole-anchored sampling; an inner level, via motion-driven attention, localizes the blood-mass motion and carries it across the interface during filling and ejection; a probe level splits the outer transport, calibrated by the inter-frame displacement field, along its dimensional gradient into a four-path decomposition (displacement, impact, transport, burst), one head each. On EchoNet-Dynamic the model reads \(\text{R}^2=0.8082\pm0.0007\) from a single view, notably \(0.8187\pm0.0019\) with six-view test-time augmentation, and \(0.8223\) over three seeds, from EF labels and ED/ES indices alone --- no contours, masks, or volume annotations --- with the VideoMAE (Kinetics-400) backbone frozen and only the 9.97M-parameter ISNF head trained on a single RTX 5090 Laptop GPU. The accuracy is attributable: strikingly, the displacement head alone reaches \(0.809\), on par with the fused model, whereas a non-learned optical-flow probe over the same windows reaches \(0.441\) --- the readout lives in the learned, interface-respecting representation.
	
\end{abstract}

\begin{IEEEkeywords}
	Echocardiography, ejection fraction, motion-driven regression, LV-region attention, optimal transport, Neural ODE readout.
\end{IEEEkeywords}

\section{Introduction}
\label{sec:intro}

\IEEEPARstart{C}{ARDIOVASCULAR} disease leads the world in mortality and disability, and echocardiography is the first-line modality for measuring ejection fraction (EF). The conventional protocol delineates the cavity twice, at end-diastole (ED) and at end-systole (ES), and takes the ratio of the two volumes; deep learning instead regresses EF from millions of fragmented pixels, at the cost of large labelled cohorts and heavy compute. Both cut the blood--wall interface: one into two static contours, the other into an unstructured cloud. The phase trajectory along which blood and wall act on each other is discarded, and what is cut away must be relearned from data, as one more sample for millions of parameters to fit and one more epoch of compute to afford. Read as a contact surface tracked in motion the readout is well posed; read as unlinked fragments EF becomes a mapping the network must search for and piece together inside a black box under large-data supervision. We argue EF is not mapped from fragments but read out from the intact interface, one scalar condensing the heart's full cycle, from wall thickening and cavity deformation to the loop closing from ED back to ED.

The fraction is a continuous, pulsatile integral, not the quotient of two static snapshots: the ventricle fills and ejects across the interface, and every instant leaves its trace there. Volumes are integrals of that boundary motion, so EF is identifiable from one anchored cycle rather than from two endpoint delineations --- carried to a scalar while staying on the interface that carries it.

Therefore, reading that contraction out is a transport at three levels. The \textit{outer} level drops a high-dimensional distribution over the motion onto a one-dimensional readout; the \textit{inner} level is that motion itself, blood mass redistributed across the interface as the ventricle fills and ejects; the \textit{probe} level splits the drop, calibrated by the inter-frame displacement field, into four paths --- displacement, impact, transport, burst --- each head a microscopic interface where one path becomes the scalar, so the four share the compression rather than one absorbing it whole. The levels meet at the interface: the outer transport is faithful only while the inner one stays exposed there, so speckle, flow noise, artefact and the chamber boundary are the mixed signal to carry, not variation to pool away. Anchoring at end-diastole closes the cycle: the same redistribution fills before the anchor and ejects after it, and must be read as one motion.

Specifically, two conditions make that anchor work. \textbf{Temporal anchoring}: the sampling follows the cardiac clock, so that one inter-frame displacement is read at one phase only. \textbf{Spatial grounding}: the readout attends to the left-ventricular (LV) cavity and interface where the mass moves, rather than to background. Without the first the inner transport is ambiguous in time; without the second it is lost in background --- in time by ED-anchored cycle sampling fixing the phase axis, and in space by LV-region attention pinning the integration boundary $\Gamma(t)$.

Specifically, existing networks treat echocardiogram videos as generic frame sequences, and the two prevailing pipelines cut the interface twice over: (i)~\emph{Interface-blind aggregation} cuts the boundary into a pooled weight, and (ii)~\emph{Flow-blind temporal modeling} cuts the motion between frames, and (iii)~\emph{Phase-uniform sampling} cuts the cycle out of time. All three serve the endpoint score rather than the physiology.

The readout restores what the cuts remove by keeping the interface intact where the three cuts open it: the sampling follows the cardiac clock rather than uniform frames, the displacement between frames is kept as the calibration of the drop instead of being pooled away, and the interface itself is exposed rather than predicted --- the left ventricle is read where the mass actually moves.

In this work, the proposed ISNF-GRU-ODE architecture reads the intact interface in three stages: an implicit spatial neural flow (ISNF) encoder for the LV interface, a bidirectional gated recurrent unit (GRU) scan along the cardiac clock, and a continuous ordinary differential equation (ODE) readout onto a single scalar. It never predicts ventricular volumes (EDV, ESV) or a contour; it exposes the interface itself --- the LV region map, the motion descriptors, and the four paths --- and checks each against the inter-frame displacement field (region map against LV anatomy, descriptors against cardiac phase, each path against EF), so the estimator is a measurement instrument whose internals are checked rather than asserted. To the best of our knowledge, the specific combination of a Neural ODE readout integrated over an ED-anchored cardiac cycle at a physical blood-wall interface, calibrated by the inter-frame displacement field and bifurcated into four paths without predicting volumes or contours, has not been reported. Our contributions are four:

\begin{itemize}
	\item \textbf{The heartbeat cycle is read by its phase, not by uniform frames.}
	The sampling window starts at end-diastole and returns there at cycle end, so one inter-frame displacement can never mean filling and ejection at once. A bidirectional GRU scan with a continuous ODE readout then carries the per-interval observations along the cardiac clock; without this the reading is ill-posed, not merely less accurate
	(Sec.~\ref{sec:method}).
	
	\item \textbf{The LV region interface is read where the blood mass moves, without a contour label.}
	A two-layer spatial attention pool produces the interface, trained through the EF and motion heads only. Speckle, flow noise and artefact stay in the signal: the LV region map marks the surface where they become legible, so the redistribution is measured rather than assumed
	(Sec.~\ref{sec:method}).
	
	\item \textbf{The compressed outer transport drop is split into four paths instead of one latent.}
	The ODE's high-dimensional state is not itself checkable, so the reading is taken where the high-dimensional motion comes down to one scalar: across one cardiac cycle the blood mass is carried, pushed, released and let through the valve, and the inter-frame displacement field calibrates that drop and splits it into four channels --- how far the wall moves (\emph{displacement}), how the mass is redistributed (\emph{transport}), when the redistribution breaks (\emph{burst}), and how much crosses the surface (\emph{impact}) --- each with its own EF head. The split is falsifiable: a path with no EF information fails its own head, a redundant one does not survive pruning. Path-only $R^{2}$ reaches $0.809$ (displacement), $0.804$ (impact), $0.794$ (transport); burst is weak alone ($0.241$) yet non-redundant, since pruning it costs $-0.0048$ $R^{2}$ and pruning burst and transport $-0.0058$ against a $0.0019$ seed spread. A four-weight bank of the path heads fitted on validation recovers $99.8\%$ of fused accuracy ($\Delta R^{2}=-0.0005$, 95\% CI $[-0.0025,+0.0015]$); a probe on the fused motion features reaches $R^{2}=0.8088$ against the full model's $0.809$, where a non-learned Farneb\"ack probe on the same windows reaches $0.441$
	(Sec.~\ref{sec:validity}).
	
	\item \textbf{Each claim is checked before the reading is deployed.}
	What the instrument reports is what the displacement field confirms: the map lines up with the left-ventricular cavity without segmentation supervision, the motion descriptors concentrate in systole, and each path decodes EF on its own head. The gain is attributed to the transport rather than capacity --- a $\pm3$-frame window shift costs $0.033$ R$^{2}$ (about $0.14$ EF points MAE per frame), cycle anchoring adds $+0.031$ R$^{2}$ with the phase aligned --- and the reported accuracy is a three-seed ensemble under test-time augmentation
	(Sec.~\ref{sec:ablation}). A learned ED/ES detector then closes the anchoring loop (Sec.~\ref{sec:deploy}); the reading transfers to adult CAMUS and pediatric EchoNet-Peds under split-conformal intervals, and a four-weight linear stack of the path heads recovers $99.8\%$ of how the fused model reads
	(Sec.~\ref{sec:main_results},
	Sec.~\ref{sec:generalization}).
\end{itemize}

\section{Related Work}
\label{sec:related}

Early deep left-ventricular EF (LVEF) work built on segmentation: a network delineates the LV~\cite{leclerc2019deep}, and volumes are computed from the masks~\cite{li2023echoefnet}. These methods depend on accurate ED/ES selection~\cite{ouyang2020video}, so frame segmentation errors propagate to EF. End-to-end regression~\cite{ouyang2020video,muhtaseb2022echocort} avoids that propagation but treats the video as a generic sequence: uniform sampling in time, global pooling in space, and frame-based operators along the way --- three practices that cut the blood-wall interface where the ejection information lives. We keep end-to-end regression and replace the three cuts with one transport: the inner blood-mass redistribution is carried along the cardiac clock by a continuous ODE over the anchored cycle and lands on the readout as a single scalar.

What EF carries is the dynamic process linking EDV and ESV, not the ratio itself. Phase-contrast magnetic resonance encodes blood velocity in spin phase, a lineage rooted in Hahn's spin-echo detection of motion~\cite{hahn1960detection,vonschulthess1985blood}. In ultrasound the analogous observable is speckle motion, estimated by optical-flow methods~\cite{horn1981opticalflow,beauchemin1995opticalflow}, and speckle-tracking strain imaging~\cite{dhooge2000strainrate,amundsen2006noninvasive} turns the same displacement measurement onto myocardial deformation. Learned ultrasound-flow networks train displacement against synthetic or tracked targets, or use cycle consistency without EF labels. Against that lineage we read the blood-side inter-frame displacement as the inner transport at the feature level rather than estimated as pixel-level optical flow, and train the four paths end-to-end against EF rather than against a flow ground truth --- EF-supervised descriptors of that transport, not a learned optical-flow sensor. A non-learned Farneb\"ack probe reaches only $\text{R}^2=0.441$ where the learned path head reaches $0.809$, locating the information in the learned, interface-respecting representation (Sec.~\ref{sec:validity}).

For temporal modeling, state-space models such as Mamba provide linear-time long-range context~\cite{gu2023mamba,ma2024umamba,liu2024vmamba}, while Neural ODEs introduce continuous-time dynamics into the network~\cite{chen2018neural}. We take the latter: a bidirectional GRU scan forms a phase-averaged cycle state, and a Neural ODE closes the outer transport over the anchored cycle --- matching the cycle-level nature of EF, whereas uniform frame-based operators would reintroduce the phase-uniform sampling and flow-blind temporal modeling this design sets out to replace.

\begin{figure*}[!t]
	\centering
	\includegraphics[width=0.76\textwidth]{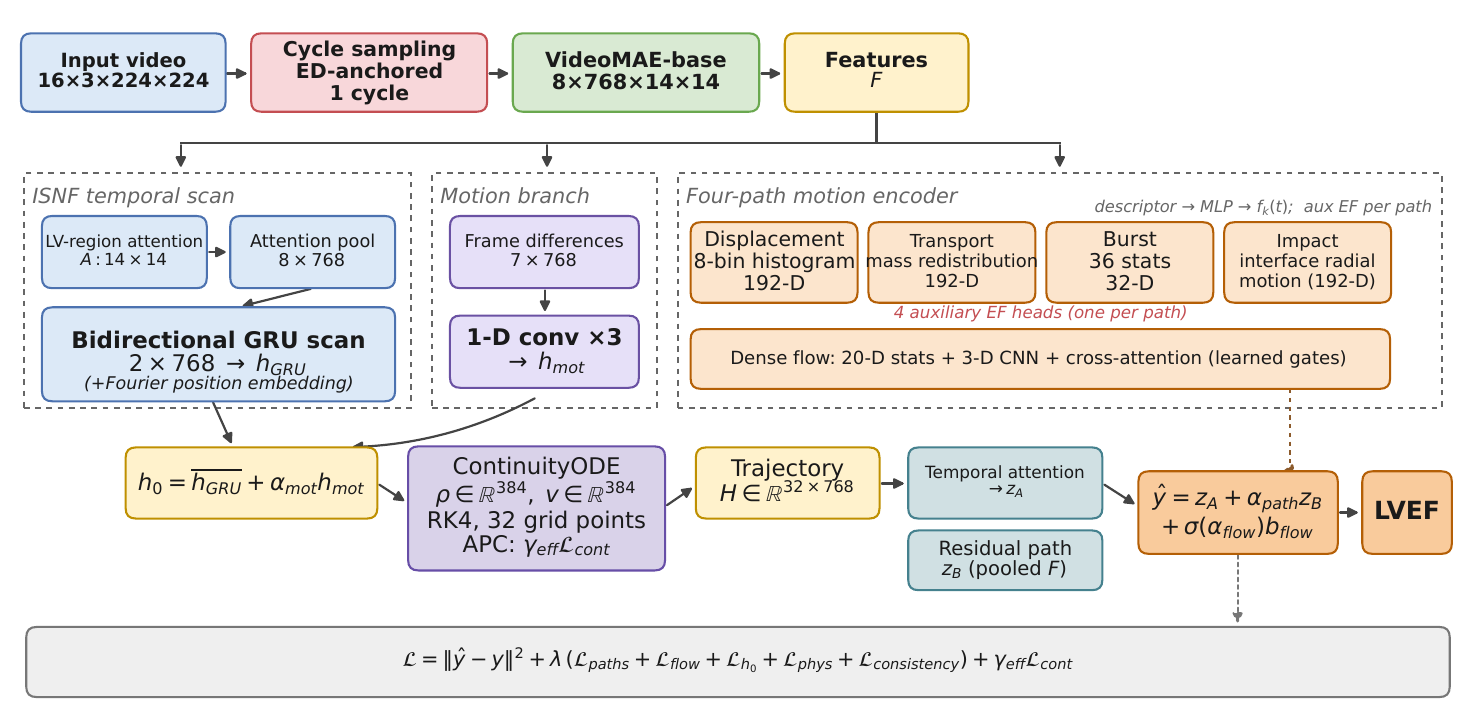}
	\caption{The three transport levels: the outer transport is the whole left-to-right chain from the anchored window to $\hat y$; the inner transport is the blood-mass redistribution localised by LV-region attention $A$ at the interface $\Gamma(t)$; the probe level is the four-path block --- the four terms of the flux integral Eq.~\eqref{eq:sv_integral}, not four sensors --- checked one channel at a time. $8\times768\times14\times14$ VideoMAE-base features from the anchored 16-frame window are pooled by LV-region attention, scanned by a bidirectional GRU, and expanded by a ContinuityODE into a 32-step RK4 readout; a temporal-attention readout $z_A$ and a residual path $z_B$ on the pooled backbone features then combine into the EF prediction. The dashed blocks (trainable ISNF head, 9.97M parameters) sit on a frozen VideoMAE-base backbone (86.23M), 96.20M in total.}
	\label{fig:pipeline}
\end{figure*}

The same compression has a mathematical backbone on the spatial side. Optimal transport maps one probability distribution onto another by the minimal-cost plan, and the 2-Wasserstein gap quantifies the mismatch between them~\cite{villani2009optimal}; its discrete special case is the classical assignment problem~\cite{munkres1957algorithms}, Sinkhorn's entropic iterations make the gap differentiable at scale~\cite{cuturi2013sinkhorn}, and Wasserstein transport has been used to compare mass distributions on geometric domains~\cite{solomon2015convolutional}. The temporal axis asks for the same continuous treatment, since Neural ODEs model continuous latent dynamics in irregular time series~\cite{rubanova2019latent}. No formulation we are aware of couples entropic-transport compression with a continuous ODE readout to carry a spatiotemporal motion distribution, across a physical blood-wall interface, onto one clinical scalar.

\section{Method}
\label{sec:method}

Fig.~\ref{fig:pipeline} gives an overview of the proposed ISNF-GRU-ODE architecture: the outer transport is the full chain from the anchored window to $\hat y$, the inner transport is the blood-mass redistribution localised at the interface by the LV-region attention, and the probe level is the four-path block that splits the inter-frame displacement field and checks it one head at a time --- the four paths corresponding one-to-one to the terms of the flux integral Eq.~\eqref{eq:sv_integral} (displacement, impact, transport, burst), so the diagram and the formulation carry the same decomposition, read left to right as phase axis (Sec.~\ref{sec:flowformulation}), interface and split (Sec.~\ref{sec:hemo}), continuous readout (Sec.~\ref{sec:isnf}). The ED-anchored 16-frame window is encoded by VideoMAE-base, and three parallel blocks read the resulting features: an LV-region attention with a bidirectional GRU scan, a motion branch on inter-frame feature differences, and a four-path encoder that splits one inter-frame displacement field into four paths inside the probe-level interface. The GRU state and the motion branch merge into the initial state $h_0$, which a ContinuityODE integrates over the anchored cycle; a softmax readout and a residual path then combine with a gated dense-flow correction into the EF prediction. Two design choices drive this shape: the four paths exist because a single pooled readout would force one scalar to carry four distinct physical quantities, and the ODE readout closes the outer transport continuously over the anchored cycle, so the scalar is an integral of the motion rather than a sum over discrete frames.

\subsection{Motion Formulation and Cycle Sampling}
\label{sec:flowformulation}
The estimator builds on two established building blocks --- VideoMAE feature extraction and GRU-based temporal scanning --- and introduces three new components: the four-path motion encoder, the ContinuityODE, and the transparent path-head bank. Reading EF from motion requires exposing the three transports of Sec.~\ref{sec:intro} in a form the network can learn. Given a video $x \in \mathbb{R}^{16 \times 3 \times 224 \times 224}$, we predict the EF fraction $y \in [0,1]$. Let $\mathbf{u}(t,\mathbf{x})$ be the instantaneous displacement field (the ultrasound analogue of optical flow), $\Gamma(t)$ the endocardial interface with normal \(\mathbf{n}(t,\mathbf{x})\), and \(\rho(t,\mathbf{x})\) the blood-mass field. Eq.~\eqref{eq:sv_integral} makes the inner transport explicit as a measurable flux across $\Gamma(t)$; the outer transport --- compressing the full motion distribution onto one scalar --- is what this section instantiates. Stroke volume (SV) is the flux across that interface,
\begin{equation}
	\text{SV} = \int_{t_{\text{sys}}} \int_{\Gamma(t)} \rho(t,\mathbf{x})\, \mathbf{u}(t,\mathbf{x}) \cdot \mathbf{n}(t,\mathbf{x})\; d\Gamma\, dt,
	\label{eq:sv_integral}
\end{equation}
so that the $\text{LVEF}=100\cdot\text{SV}/\text{EDV}$ (in percent). The equation is not evaluated numerically, since EDV is not delineated at test time; it specifies which motion quantities the encoder must expose ---$\mathbf{u}$, the redistribution of $\rho$, and the radial motion $\mathbf{u}\cdot\mathbf{n}$ --- and the four-path encoder below provides discrete surrogates: displacement and impact paths for $\mathbf{u}$ and $\mathbf{u}\cdot\mathbf{n}$, transport and burst for the redistribution of $\rho$.

Cycle sampling requires ED/ES indices: training uses only the annotated ED/ES frames of each training video, with the EF label as the sole regression target and no contour, segmentation or volume annotations; at test time the indices come either from the reference tracings (comparison protocol) or from a learned detector (deployment protocol, Sec.~\ref{sec:deploy}). With $\Delta=2(\text{ES}-\text{ED})$ estimating cycle length, we sample $T=16$ indices
\begin{equation}
	\text{idx}(t) = \left\lfloor \text{ED} + \frac{t}{T-1}(\Delta-1) \right\rfloor,
	\label{eq:cycle_idx}
\end{equation}
so that $t=0$ is ED and the window spans about one cycle: within one anchored cycle the same displacement means filling before the anchor and ejection after it, so no interval is ambiguous. Overflow is wrapped modulo video length so that frame~0 remains exactly ED; otherwise the loader falls back to uniform sampling.

\subsection{Four-Path Motion Encoder}
\label{sec:hemo}
The four-path motion encoder exposes four complementary paths, one per term of the flux integral: the displacement field $\mathbf{u}$ itself (displacement), the radial component $\mathbf{u}\cdot\mathbf{n}$ at $\Gamma(t)$ (impact), the redistribution of blood mass $\rho$ (transport), and its time-localised transient (burst). They are four faces of one inter-frame displacement field, not four sensors, all computed from the same field and the LV-region attention map at the calibrated probe-level interface.\label{app:F1}

\textbf{Displacement (systolic motion intensity).} Per-interval feature differences are projected onto a 2-D displacement vector and accumulated into a soft 8-bin orientation histogram over the attention-weighted blood pool, weighted by $\max(0,\cos\Delta\theta)$; mean displacement, anisotropy-weighted magnitude and the histogram form the descriptor, mapped by a two-layer MLP to $f_{\text{dir}}(t)\in\mathbb{R}^{192}$. 

\textbf{Transport (mass redistribution).} Attention-weighted mass fields on a $14\times14$ grid are transported between intervals by entropic optimal transport~\cite{cuturi2013sinkhorn} (Sinkhorn, $\varepsilon=0.05$, 20 iterations); six observables --- transport cost, transported diagonal mass, its entropy, centroid shift, $L_1$ redistribution, and the signed rate of the radius of gyration --- give 19 statistics mapped to $f_{\text{ot}}\in\mathbb{R}^{192}$. This is the only path owning a mass field and a conservation residual, whose marginal error is $\mathcal{L}_{\text{cont}}$ of Eq.~\eqref{eq:apc}; its curves also feed the burst and impact paths. By a \emph{compressed transport readout} we mean that the four paths together span the EF information of the motion layer --- not that the transport path outscores the others, nor that it carries the largest weight in a near-collinear bank (Sec.~\ref{sec:validity}).

\textbf{Burst (ejection transient).} Derivatives of three transport curves (cost, centroid shift, $L_1$ redistribution) are summarized by peak height, time, width, areas and asymmetry plus cycle-level statistics; a two-layer MLP maps 36 statistics to $f_{\text{burst}}\in\mathbb{R}^{32}$.

\textbf{Impact (interface radial motion).} Inter-frame displacement is decomposed along the interface normal, estimated from the attention-map gradient, into radial and tangential components; a centroid-anchored radius-of-gyration rate supplies a signed radial flux --- the discrete counterpart of $\mathbf{u}\cdot\mathbf{n}$ in Eq.~\eqref{eq:sv_integral}, i.e. the inner transport across the interface, exposed rather than predicted as a contour. A nine-statistic descriptor maps to $f_{\text{imp}}(t)\in\mathbb{R}^{192}$.

\textbf{Dense-flow conditioning.} Precomputed Farneb\"ack flow (15 fields, $64\times64$) enters as a 20-D statistic vector added to the fused representation and as six gated dimensions on transport observables, with a lightweight 3-D convolution encoding the dense field into a global descriptor and a $4\times4$ token grid queried by cross-attention, all gates learned.

\textbf{Independent heads and fusion.} Each path has an independent linear EF head,
\begin{equation}
	\hat{y}_k = \text{Linear}_k\!\left(\frac{1}{T}\sum_t f_k(t)\right),\quad k\in\{\text{dir, ot, burst, imp}\},
	\label{eq:aux_ef}
\end{equation}
trained with an auxiliary MSE so that each path is read out on its own head and becomes inspectable rather than asserted. The four streams (192+192+32+192) are fused by a two-layer MLP (608$\to$192$\to$192), and content-conditioned temporal attention aggregates the interval tokens into $z$, where two linear probes give $\hat y_{\text{flow}}$ and $\hat y_{\text{phys}}$, both trained against $y$ and each a probe-level interface (Sec.~\ref{sec:validity}); only the gated $\mathbf{b}_{\text{flow}}$ enters Eq.~\eqref{eq:flow_bias}:
\begin{equation}
	\hat{y}=z_A+\alpha_{\text{path}}z_B+\sigma(\alpha_{\text{flow}})\cdot\mathbf{b}_{\text{flow}},
	\label{eq:flow_bias}
\end{equation}
where $\alpha_{\text{path}}=0.056$ and $\alpha_{\text{flow}}$ is learnable.

\subsection{Temporal Module and Training}
\label{sec:isnf}
\textbf{LV-region attention.} A two-layer spatial attention pool produces $A=\text{softmax}_{H'W'}(\text{Conv3d}_{1\times1}(F_{\text{feat}}))$, and $F_{\text{pool}}(t)=\sum_{h,w}A(\cdot,t,h,w)F_{\text{feat}}(\cdot,t,h,w)$. Trained only through EF and the motion heads, $A$ localizes the cavity and interface --- the site of the inner transport --- without segmentation, satisfying the spatial-grounding prerequisite, with a zero-initialized Fourier embedding of temporal position added before scanning.

\textbf{Scan and motion.} A bidirectional GRU scans $F_{\text{pool}}$ in both directions and concatenates the hidden states; in parallel, spatially pooled inter-frame feature differences are encoded by a three-layer 1-D convolution over the seven feature-step intervals into $h_{\text{mot}}$, with $h_0=\bar h_{\text{GRU}}+\alpha_{\text{mot}}h_{\text{mot}}$ ($\alpha_{\text{mot}}$ initialized at 0.1). A linear head reads $\hat y_{h_0}$ from $h_0$ as deep supervision before integration.

\textbf{Continuity ODE.} The state splits into mass $\rho\in\mathbb{R}^{384}$ and momentum $v\in\mathbb{R}^{384}$, integrated by a fixed-step fourth-order Runge-Kutta (RK4) scheme over 32 grid points:
\begin{equation}
	\begin{aligned}
		\dot{\rho} &= -\big(\mathrm{softplus}(\rho)+10^{-6}\big)\odot\tanh(W_{\mathrm{div}}v),\\
		\dot{v} &= \mathrm{MLP}([\rho,v,t]),
	\end{aligned}
	\label{eq:ode}
\end{equation}
with $W_{\mathrm{div}}$ initialized as two identity blocks and the momentum MLP as a small-initialized residual (769$\to$192$\to$384). The continuity structure closes the outer transport: $\dot\rho$ is a discrete mass-conservation law, so the readout integrates flux along the cycle rather than pooling frames, matching Eq.~\eqref{eq:sv_integral}. A softmax temporal readout aggregates the trajectory into $z_A$, with a residual path on the spatially pooled backbone features supplying $z_B$.

\textbf{Adaptive physics coupling (APC).} The continuity residual $\mathcal{L}_{\text{cont}}$ is the normalized marginal error of the Sinkhorn plan, averaged over grid and cycle, with learned weight:
\begin{equation}
	\mathcal{L}_{\text{apc}}=\gamma_{\text{eff}}\mathcal{L}_{\text{cont}},\qquad
	\gamma_{\text{eff}}=\exp(\ell_\gamma)\,\sigma\!\left(\beta-\sqrt{\mathcal{L}_{\text{cont}}+\epsilon}\right),
	\label{eq:apc}
\end{equation}
with $\epsilon=10^{-8}$ and $\exp(\ell_\gamma)$ initialized to 0.1; when the Sinkhorn marginal error is small, $\gamma_{\text{eff}}$ increases, enforcing the conservation constraint more strongly where it is already nearly satisfied.

\textbf{Objective.} The total loss is
\begin{equation}
	\begin{aligned}
		\mathcal{L}={}&\|\hat y-y\|^2+\lambda\Big(\sum_k\|\hat y_k-y\|^2+\|\hat y_{\text{flow}}-y\|^2\\
		&+\|\hat y_{h_0}-y\|^2+\|\hat y_{\text{phys}}-y\|^2+\sum_k\|\hat y_k-\mathrm{sg}(\hat y)\|^2\Big)\\
		&+\gamma_{\text{eff}}\mathcal{L}_{\text{cont}},
	\end{aligned}
	\label{eq:total_loss}
\end{equation}
with all auxiliary weights $\lambda=0.1$; $\mathrm{sg}(\cdot)$ stops the fused prediction from being pulled by the auxiliary heads in the consistency term.

\section{Experiments}
\label{sec:experiments}
\noindent\textit{Reading convention for this section.}
Throughout, \emph{outer transport} denotes the distribution-to-scalar map that carries one anchored heartbeat cycle onto the EF reading, \emph{inner transport} the blood-mass redistribution across the blood-wall interface, and \emph{probe-level transport} that same map restricted to one channel, where a probe (a linear readout on a fixed feature set; the \emph{non-learned reference probe} (Farneb\"ack optical flow) is a probe on a fixed, untrained feature set) or path head realizes the interface and the channel's own accuracy measures the transmission. The transport path --- one of the four paths, computed via Sinkhorn optimal transport --- is a distinct object from the outer transport: the former measures mass redistribution at the probe level, the latter is the architectural function of the whole readout chain. The flux integral in Eq.~\eqref{eq:sv_integral} carries the first two senses; the probe level is not a third sense of it but the third place where the same two are checked, one channel at a time. The three mechanisms under test are the closed-loop cycle sampler (the phase axis of the outer transport), the LV-region attention map (which isolates the inner transport), and the four-path encoder (four faces of one inter-frame displacement field). The first two mechanisms implement the two prerequisites of Sec.~\ref{sec:intro} --- temporal anchoring and spatial grounding; the third makes the factorisation testable at the probe level.
\label{app:X0}

\subsection{Datasets and Implementation}
\label{sec:setup}
We evaluate on \textbf{EchoNet-Dynamic}~\cite{ouyang2020video}: 10{,}030 apical-4-chamber videos split into 7{,}465 training, 1{,}288 validation, and 1{,}277 test videos; 10{,}025 carry ED/ES tracings, used for cycle sampling at training and as the oracle-anchor comparison protocol at test; the deployment protocol replaces them with detector predictions (Sec.~\ref{sec:deploy}). Inputs are $224\times224$ with $T=16$ frames. Models are trained for 60 epochs with batch size 12 and AdamW (learning rate $10^{-4}$, weight decay 0.05), 3-epoch warmup, cosine decay, exponential moving average (EMA) 0.999, and a VideoMAE-base backbone~\cite{tong2022videomae} fine-tuned with layer-wise learning-rate scale 0.3 (decay 0.85); the ODE width is 192. We report $\text{R}^2$, mean absolute error (MAE), and root-mean-square error (RMSE); TTA averages six views (uniform, $\pm2$-frame offset, horizontal flip). All experiments run on one RTX 5090 Laptop GPU.

\subsection{Main Results and Comparison}
\label{sec:main_results}
Across seeds 42--44, the final model reaches $\text{R}^2=0.8082\pm0.0007$ single-view (MAE $4.05\pm0.02\%$) and $0.8187\pm0.0019$ with six-view TTA (MAE $3.96\pm0.02\%$); a 3-seed TTA ensemble reaches $0.8223$ (MAE $3.91\%$), with $\pm$ denoting the sample standard deviation over the three seeds. Table~\ref{tab:main} summarizes the three-seed result and Table~\ref{tab:sota} compares methods from 2018--2026: on the official split the best reported value (Echo-Vision-FM~\cite{zhang2025echovisionfm}, $\text{R}^2=0.825$) is matched by our ensemble within $0.003$.

These numbers come from the three mechanisms, not from the backbone alone. The cycle sampler supplies the phase axis of the outer transport; the LV-region map isolates the inner transport; the four paths expose the displacement field from four directions. The scalar is transported from a phase-aligned cycle instead of pooled from generic frames.
\label{app:X1}

\begin{table}[!htbp]
\centering
\begin{threeparttable}
\caption{ISNF-GRU-ODE on EchoNet-Dynamic (test, $n=1{,}277$). Single-view and 6-view TTA are mean $\pm$ standard deviation over seeds 42--44; the ensemble averages the three TTA predictions.}
\label{tab:main}
\footnotesize
\setlength{\tabcolsep}{4pt}
\begin{tabular}{lccc}
\toprule
Metric & Single view & 6-view TTA & Ensemble \\
\midrule
R$^2$ $\uparrow$ & $0.8082 \pm 0.0007$ & $0.8187 \pm 0.0019$ & \textbf{0.8223} \\
MAE (\%) $\downarrow$ & $4.05 \pm 0.02$ & $3.96 \pm 0.02$ & \textbf{3.91} \\
RMSE (\%) $\downarrow$ & $5.35 \pm 0.01$ & $5.21 \pm 0.03$ & \textbf{5.15} \\
\bottomrule
\end{tabular}
\begin{tablenotes}[flushleft]
\item Seed-42 bootstrap 95\% confidence interval (CI): R$^2$ $[0.782, 0.830]$, MAE $[3.84, 4.22]$, RMSE $[5.11, 5.60]$. Giving each view its own cached flow field changes the seed-42 single-view R$^2$ from $0.809$ to $0.8093$.
\end{tablenotes}
\end{threeparttable}
\end{table}
\begin{table}[!htbp]
\centering
\begin{threeparttable}
\caption{Comparison with state-of-the-art methods on EchoNet-Dynamic. \emph{Official} = EchoNet-Dynamic split; \emph{custom} = self-made split, not directly comparable. Ours: 3-seed 6-view TTA ensemble.}
\label{tab:sota}
\footnotesize
\setlength{\tabcolsep}{4pt}
\begin{tabular}{lcccc}
\toprule
Method & Year & Split & R$^2$ $\uparrow$ & MAE (\%) $\downarrow$ \\
\midrule
MC3~\cite{tran2018closer} & 2018 & official & 0.76 & 4.45 \\
R3D~\cite{hara2018can} & 2018 & official & 0.79 & 4.22 \\
EchoNet-Dynamic~\cite{ouyang2020video} & 2020 & official & 0.81 & 4.10 \\
UVT~\cite{reynaud2021ultrasound} & 2021 & custom\tnote{a} & 0.52 & 5.95 \\
EchoCoTr~\cite{muhtaseb2022echocort} & 2022 & official & 0.82 & 3.95 \\
Hierarchical ViT~\cite{fazry2022hierarchical} & 2022 & official & 0.59 & 5.59 \\
EchoGraphs~\cite{thomas2022echographs} & 2022 & custom\tnote{a} & 0.81 & 4.01 \\
ViViEchoformer~\cite{akan2024viviechoformer} & 2025 & official & 0.55 & 6.14 \\
EFNet~\cite{ali2025efnet} & 2025 & custom\tnote{a} & 0.82 & 3.70 \\
Echo-E$^3$Net~\cite{heidari2025echoe3net} & 2025 & official & 0.82 & 3.92\tnote{b} \\
SGS EchoNet~\cite{artman2025sgs} & 2025 & custom\tnote{c} & 0.56 & 5.73 \\
Echo-Vision-FM~\cite{zhang2025echovisionfm} & \red{2025} & official & 0.825 & 3.87 \\
\midrule
\textbf{ISNF-GRU-ODE (ours)} & 2026 & official & 0.8223 & 3.91 \\
\bottomrule
\end{tabular}
\begin{tablenotes}[flushleft]
\item MC3 and R3D are benchmarked by Ouyang et al.~\cite{ouyang2020video}.
\item[a] Self-made random splits (75/12.5/12.5 for UVT, 80/10/10 for EchoGraphs and EFNet); EchoGraphs reports R$^2=0.81$ for its 16-frame single-heartbeat regression head.
\item[b] Echo-E$^3$Net: RMSE $=5.20$.
\item[c] SGS: authors' Stanford split; R$^2$ computed as $r^2$ from the reported $r=0.75$.
\end{tablenotes}
\end{threeparttable}
\end{table}

\subsection{Ablation}
\label{sec:ablation}
Table~\ref{tab:building} isolates the components. Cycle anchoring adds $+0.031$ $\text{R}^2$ over uniform sampling and the ISNF temporal model another $+0.031$ --- the two prerequisites carry the accuracy. The four-path motion encoder and APC together bring TTA to $0.820$, statistically indistinguishable from the temporal model alone, so the probe level carries the check (Sec.~\ref{sec:validity}). Readout pruning from the full model quantifies each path's marginal contribution: removing the burst readout (three paths) gives TTA $0.8156$ ($-0.0048$), and removing burst and transport readouts (two paths) $0.8146$ ($-0.0058$) --- drops of about $0.005$, consistent with a small but nonzero marginal contribution of each readout group, while the bulk of the accuracy rests on anchoring and temporal modeling. The two prerequisites therefore set the accuracy floor, while the probe level sets the verifiability ceiling; the ablation shows that neither can be removed without degrading the other.

Read this way, the decomposition follows the two prerequisites in order: cycle sampling fixes the phase axis of the outer transport, the temporal model carries it onto the readout, the regional prior isolates the inner transport, and the per-path pruning numbers measure what each face adds once that axis is well posed.
\label{app:X2}

\begin{table}[!t]
\centering
\begin{threeparttable}
\caption{Building-up ablation on EchoNet-Dynamic (seed 42; test $n=1{,}277$). All rows share the VideoMAE backbone and 60-epoch protocol; TTA is 6-view.}
\label{tab:building}
\footnotesize
\setlength{\tabcolsep}{2pt}
\begin{tabular}{lcccc}
\toprule
Configuration & Test R$^2$ & MAE (\%) & R$^2$ (TTA) & $\Delta$ (TTA) \\
\midrule
1. VideoMAE + mean-pool & 0.751 & 4.63 & n/a & -- \\
2. + Cycle sampling (ED anchor) & 0.782 & 4.27 & n/a & -- \\
3. + ISNF temporal & 0.813 & 3.99 & 0.819 & -- \\
4. + Four-path motion encoder & 0.801 & 4.11 & 0.817 & $-$0.002 \\
5. + Adaptive coupling (full) & \textbf{0.809} & \textbf{4.02} & \textbf{0.8204} & +0.003 \\
\midrule
Control: no flow/APC & 0.804 & 4.08 & 0.818 & $-$0.002 \\
\midrule
Prune burst readout & 0.8071 & -- & 0.8156 & $-$0.0048 \\
Prune burst+transport & 0.8057 & -- & 0.8146 & $-$0.0058 \\
\bottomrule
\end{tabular}
\begin{tablenotes}[flushleft]
\item Step 3 is GRU+ODE+attention; step 5 adds adaptive physics coupling. Step 4 replaces the fused readout with the four-path readout, so the single-view column is not monotonic; the full configuration is the best under TTA. Pruning rows fine-tune 3 epochs from step 5 with the burst readout (3 paths left) or burst+transport readouts (2 paths left) removed. The full configuration over three seeds gives $0.8082 \pm 0.0007$ single-view and $0.8187 \pm 0.0019$ TTA. Adjacent rows are compared pairwise on identical test videos (video-level bootstrap, $B{=}20{,}000$, seed 42): the four-path readout lowers single-view $\text{R}^2$ by $0.0119$ ($[0.0038,0.0202]$, $p{=}0.004$) and APC raises it by $0.0080$ ($[0.0026,0.0135]$, $p{=}0.004$), leaving steps 4--5 indistinguishable from step 3 (net $-0.0039$, $p{=}0.30$).
\end{tablenotes}
\end{threeparttable}
\end{table}

Inference-time interventions confirm that the reading depends on correctly localized motion, not on background texture (Table~\ref{tab:interventions}). Across all three seeds, shuffling or reversing frames costs $0.20$--$0.24$ $\text{R}^2$, freezing the input collapses the readout, removing the LV region or keeping only background is strongly destructive, and shifting the attention map by three cells lowers $\text{R}^2$ by $0.012$--$0.050$ in every seed. The estimator therefore uses the intended region, the systolic direction, and the learned attention location.

\begin{table}[!htbp]
\centering
\begin{threeparttable}
\caption{Inference-time intervention battery (test $n=1{,}277$, seeds 42--44, single view). Entries are paired $\Delta\text{R}^2$ against the matched intact input (video-level bootstrap, $B{=}20{,}000$); every 95\% CI excludes zero except window translation $\pm1$ (all seeds) and $\pm2$ at seed 44, where the $\pm2$ CI clears zero by less than $0.001$ at seeds 42--43.}
\label{tab:interventions}
\footnotesize
\setlength{\tabcolsep}{2pt}
\begin{tabular*}{\columnwidth}{@{\extracolsep{\fill}}lccc}
\toprule
Intervention & $\Delta\text{R}^2$ (s42) & (s43) & (s44) \\
\midrule
\multicolumn{4}{l}{\emph{Temporal structure}}\\
Frame shuffle & $-$0.199 & $-$0.229 & $-$0.244 \\
Frame reverse & $-$0.201 & $-$0.217 & $-$0.215 \\
Freeze first frame & $-$13.6 & $-$14.7 & $-$13.0 \\
\midrule
\multicolumn{4}{l}{\emph{Spatial content}}\\
Background only & $-$3.20 & $-$3.65 & $-$4.06 \\
LV region removed & $-$0.244 & $-$0.255 & $-$0.224 \\
Center-region crop & $-$0.061 & $-$0.066 & $-$0.056 \\
Spatial scramble & $-$0.467 & $-$0.402 & $-$0.437 \\
\midrule
\multicolumn{4}{l}{\emph{Attention location}}\\
Shift map 3 cells & $-$0.050 & $-$0.012 & $-$0.045 \\
\midrule
\multicolumn{4}{l}{\emph{Window translation (frames)}}\\
$\pm1$ & $-$0.004 & $-$0.004 & $-$0.004 \\
$\pm2$ & $-$0.011 & $-$0.011 & $-$0.008 \\
$\pm3$ & $-$0.035 & $-$0.032 & $-$0.030 \\
\bottomrule
\end{tabular*}
\begin{tablenotes}[flushleft]
\item Interventions are applied at inference only; each is compared against the same model on intact (or re-matched) inputs. Freeze collapses the ODE readout (out-of-distribution static input). Window translation shifts both anchors by the same offset, moving the sampled window without changing its length; detector anchors change start and length together and are a different intervention. Spatial scramble permutes tokens identically across frames, preserving temporal correspondence; a per-frame permutation costs $1.81$ (seed 42).
\end{tablenotes}
\end{threeparttable}
\end{table}

\subsection{Annotation-Free Deployment}
\label{sec:deploy}
The comparison protocol above anchors the cycle with the reference ED/ES indices; deployment must not require them. We close this loop with a lightweight ED/ES detector: on frozen VideoMAE features, a 64-bin frame-regression head with a confidence score slides a 24-frame window over the video and keeps the highest-confidence ED/ES hit. The detector is trained only on training-split tracings and never sees test annotations. Replacing the oracle anchors with detector predictions leaves accuracy statistically unchanged: single-view $\text{R}^2$ is $0.8044\pm0.0013$ and six-view TTA is $0.8154\pm0.0006$ across seeds 42--44, recovering $99.6\%$ of the oracle-anchored TTA ($\text{R}^2$ ratio to $0.8187\pm0.0019$) and $99.5\%$ single-view ($0.8082\pm0.0007$); the paired difference is $\Delta\text{R}^2=-0.0009$ to $-0.0055$ per seed with every 95\% confidence interval (CI) containing zero ($p>0.48$, video-level bootstrap, $B=20{,}000$ with a fixed seed).

\begin{figure*}[!t]
	\centering
	\begin{minipage}[b]{0.455\textwidth}
		\centering
		\includegraphics[width=\linewidth]{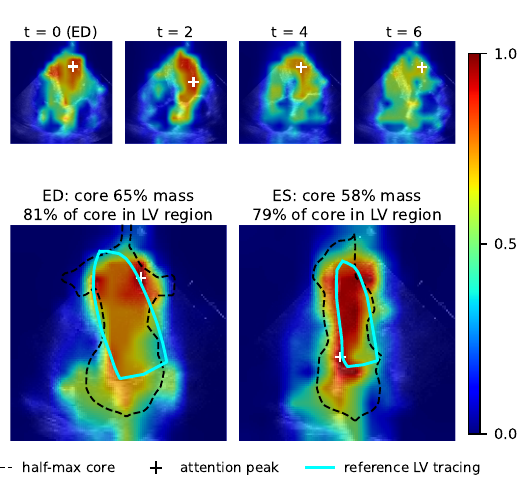}
	\end{minipage}\hfill
	\begin{minipage}[b]{0.455\textwidth}
		\centering
		\includegraphics[width=\linewidth]{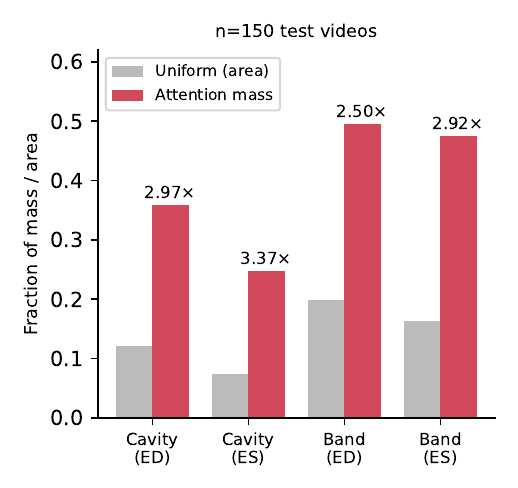}
	\end{minipage}
	\caption{Spatial grounding without supervision: the inner transport's integration boundary $\Gamma(t)$, located rather than assumed. Second prerequisite of Sec.~\ref{sec:intro} and the spatial half of contribution~2: it isolates the inner transport before any channel is split (Fig.~\ref{fig:pipeline}). Top: attention weight at four temporal tokens of one ED-anchored cycle clip; one token spans two frames, $t=0$ is end-diastole, so the panels read ED $\rightarrow$ ES $\rightarrow$ ED, + marks the peak. Bottom: ED/ES of a second median-enrichment example with the reference tracing (cyan) and the half-maximum core (dashed) --- the core falls on the traced endocardium, so the map is read along the contour. Right: over 150 test videos the half-max core holds 61\%/55\% of the mass, 83\%/76\% inside the LV region, which covers 24\%/17\% of the frame (enrichment ratios in the text). The map marks where speckle, flow noise and artefact become legible; it is not asked to denoise them away.}
	\label{fig:attn}
\end{figure*}

\begin{figure}[!t]
	\centering
	\includegraphics[width=0.95\columnwidth]{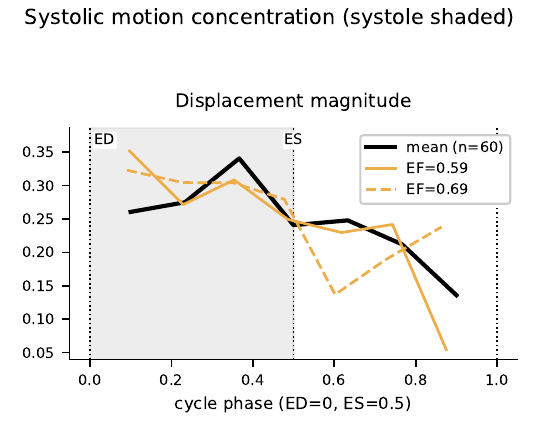}
	\caption{Temporal anchoring: the phase axis carries the outer transport. First prerequisite of Sec.~\ref{sec:intro} (temporal anchoring) and the temporal half of contribution~1: because the window is anchored at end-diastole, the same displacement cannot mean filling and ejection at once. Phase-normalised displacement magnitude (ED$=0$, ES$=0.5$; systole shaded) for the $n=60$ aggregate (black) and two example videos. Motion concentrates in systole ($+37\%$, $53/60$ videos), peaking mid-systole and falling after ES --- that shape is the cardiac-clock signature anchoring recovers, drawn from the same displacement field that calibrates the four path; attention mass inside the LV region is $0.370$ in systole versus $0.330$ in diastole, $60/60$ videos, so phase axis and the interface are read together.}
	\label{fig:curves}
\end{figure}

That anchoring tolerance prices the error of the outer transport's phase axis: a mis-anchor displaces the same blood-mass redistribution to the wrong side of the cardiac clock, exactly as the intact-interface account predicts. Because EF is invariant to which cycle is observed, the detector's coarse anchors still close the loop.
\label{app:X3}

The detector's own anchors are coarse (exact reference match in $7.8\%$ of videos, within $\pm2$ frames in $38.7\%$): what the reading requires is that a complete contraction---relaxation cycle be covered, not an absolute phase. The instrument thus runs end-to-end without manual ED/ES selection at test time.

\subsection{Verifiable Measurement}
\label{sec:validity}
\begin{table}[!t]
\centering
\begin{threeparttable}
\caption{Independent EF prediction of each motion path (test, single view, seed 42, $n=1{,}277$).}
\label{tab:fourpath}
\footnotesize
\setlength{\tabcolsep}{4pt}
\begin{tabular}{lccc}
\toprule
Path & Motion quantity & R$^2$ $\uparrow$ & MAE (\%) $\downarrow$ \\
\midrule
Displacement & Systolic motion intensity & \textbf{0.809} & 4.03 \\
Impact & Interface radial motion & 0.804 & 4.14 \\
Transport & Mass redistribution & 0.794 & 4.23 \\
Burst & Ejection transient & 0.241 & 7.88 \\
\midrule
Full model (fused) & All paths + ISNF & \textbf{0.809} & 4.02 \\
\bottomrule
\end{tabular}
\begin{tablenotes}[flushleft]
\item Each path is evaluated through its own dedicated linear EF head (Eq.~\eqref{eq:aux_ef}), independent of the fusion module. The four values are single-path ceilings, not a ranking inside the fused readout, where the four heads enter one four-weight bank. Burst is weak alone but contributes in fusion (readout pruning, Table~\ref{tab:building}). Transport is the structural rather than the strongest channel: the four path-head predictions are near-collinear ($r=0.987$--$0.995$ among displacement, transport, and impact, burst $0.55$). Transport is also the only path owning a mass field and a conservation residual, its radius-of-gyration trend being the interface channel's signed radial observable and its cost curve the input that is differentiated to form the transient channel (Sec.~\ref{sec:hemo}).
\end{tablenotes}
\end{threeparttable}
\end{table}

Fig.~\ref{fig:attn} checks the interface (inner transport), Fig.~\ref{fig:curves} checks the phase axis (outer transport), and the path heads with Fig.~\ref{fig:measurement} the probe level as accuracy and as uncertainty.

\emph{Spatial grounding.} Without segmentation supervision the LV-region attention aligns with the cavity and interface in Fig.~\ref{fig:attn}; over 150 test videos it enriches the cavity $2.97\times$ at ED and $3.37\times$ at ES, and an 8-px boundary band $2.50\times$/$2.92\times$, relative to uniform area; the overlay examples are median-enrichment cases, not selected extremes. Read along the traced endocardial contour the attention density is $2.85\pm0.55\times$ the frame average, forming one continuous arc over $66\pm31\%$ of the contour (three seeds, 1{,}276 test videos). The map is therefore not a post-hoc explanation: it is the measurement region, located before the scalar exists.

\emph{Temporal grounding.} In Fig.~\ref{fig:curves}, phase-normalised motion descriptors concentrate in systole to demonstrate the effect of temporal grounding: displacement magnitude averages $0.280$ in systole versus $0.204$ in diastole ($+37\%$), consistent in $53/60$ videos; attention mass within the LV region is higher in systole in $60/60$ videos ($0.370$ vs.\ $0.330$). Uniform sampling would smear exactly this shape; the same displacement draws the curve and calibrates the four paths.

\emph{Descriptor-level evidence.} Because each path has an independent EF head (Eq.~\eqref{eq:aux_ef}), we measure how much EF information each channel carries (Table~\ref{tab:fourpath}): path-only $\text{R}^2$ is $0.809$ for displacement, $0.804$ for impact, and $0.794$ for transport, with burst weaker alone ($0.241$) but contributing in fusion. A probe on the fused motion features reaches $\text{R}^2=0.8088$, matching the full model's seed-42 value of $0.809$ to within $0.0002$, whereas a non-learned Farneb\"ack probe reaches only $0.441$: the EF signal is carried by the learned motion representation. Each probe is the outer transport restricted to one channel's support, so its accuracy reports how much inner transport crosses that interface. The interface descriptor is repeatable across the six TTA views (intraclass correlation coefficient (ICC) $(1,1)=0.954$--$0.974$) and tracks EF across populations (Sec.~\ref{sec:generalization}). After the per-head affine calibration the transport head reads, over three seeds, $\text{R}^2=0.791\pm0.004$; a degree-2 fit on the mean values of four transport observables (transport cost, centroid shift, $L_1$ redistribution, radius-of-gyration rate), fitted on validation and evaluated on test, reads $0.744\pm0.089$ over seeds and, at seed 42, $0.796$ against the head's own $0.794$ --- the surrogate explains the head without beating it.

Taken together, these three checks are the evidence for the three transports: the enrichment locates the site of the inner transport, the systolic concentration shows the outer transport is carried by the four paths, and the gap to the Farneb\"ack probe measures the probe-level transport, identifying the EF signal with the learned transport itself.
\label{app:X4}

\emph{Transparent readout.} The four path heads can be stacked into a transparent bank,
\begin{equation}
	\hat y_{\text{bank}}=\sum_{k} w_k \hat y_k + b,
	\label{eq:bank}
\end{equation}
with four weights fitted on validation (seed 42; weights per standard deviation: displacement $0.081$, transport $0.022$, burst $-0.001$, impact $0.010$). The bank reaches $\text{R}^2=0.8065\pm0.0018$ versus $0.8082\pm0.0007$ for the fused head ($99.8\%$ recovery; $\Delta\text{R}^2=-0.0005$, 95\% CI $[-0.0025,+0.0015]$); at the deployed six-view operating point, $0.8196$ vs.\ $0.8203$ ($99.9\%$; $\Delta\text{R}^2=-0.0008$, 95\% CI $[-0.0027,+0.0012]$). \emph{Burst} enters fusion through its transients rather than a positive weight, which is why pruning it costs less while fusion still benefits; the weights partition one near-collinear signa, so the compressed-readout claim rest on a few transport quantities sufficing, not on ranking channels.

\subsection{The Three Transports in Evidence}
\label{sec:twotransports}
The four blocks above are one statement about the blood-wall interface: Eq.~\eqref{eq:sv_integral} carries the inner transport --- the flux across $\Gamma(t)$ --- and the outer transport --- the scalar it is compressed into --- while the per-path heads measure the third, probe-level sense on each channel. Each is answered by a number.
\begin{itemize}
	\item \textbf{Temporal anchoring (phase axis of the outer transport).} Anchoring adds $+0.031$ $R^{2}$; a $\pm1$-frame anchor jitter is statistically indistinguishable from the intact input (Table~\ref{tab:interventions}), and a learned detector's coarse anchors still recover $99.6\%$ of oracle-anchored TTA.
	\item \textbf{Spatial grounding (isolating the inner transport).} Trained without a contour label, the attention map enriches the cavity $2.97\times$/$3.37\times$ at ED/ES and the 8-px interface band $2.50\times$/$2.92\times$, and displacing it by three cells is destructive in every seed.
	\item \textbf{Four faces of one displacement field.} Path-only heads reach $R^{2}$ up to $0.809$, the fused-motion probe $0.8088$ against $0.441$ for a non-learned Farneb\"ack probe, and a four-weight bank recovers $99.8\%$ --- no hidden capacity carries the decision.
\end{itemize}
The intact interface is therefore not an adjective but a testable condition: keep the interface unbroken and the readout is well posed.

\subsection{Robustness, Generalization, and Clinical Readouts}
\label{sec:generalization}
\begin{figure}[!t]
	\centering
	\includegraphics[width=0.95\columnwidth]{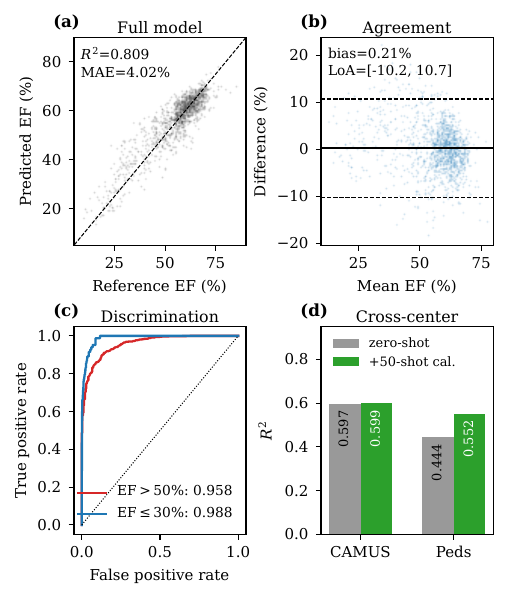}
	\caption{The probe level cashed out as a measurement performance (seed 42), reporting what the intact chain reads: the scalar at the end of the outer transport, the agreement and ranking that scalar carries, and the same reading under a changed population. Panel (a) is the outer transport end to end --- predicted versus reference EF (single view), the fused value against the reference ratio. Panel (b) is its calibration, Bland-Altman agreement: a small bias ($0.21\%$) with 95\% limits $[-10.2,10.7]\%$, i.e.\ the reading is a measurement with a stated error, not a score. Panel (c) is the same scalar as a ranking device, receiver-operating-characteristic (ROC) curves and their area under the curve (AUC); 5-fold cross-validation means are given in the text. Panel (d) is the outer transport re-anchored on a new population: held-out cross-center $R^2$, zero-shot versus 50-sample calibration --- the phase axis and the interface are re-identified rather than re-learned, which is why two scalars recover most of the gap on EchoNet-Peds while leaving CAMUS unchanged.}
	\label{fig:measurement}
\end{figure}

Under additive Gaussian noise (seed 42), TTA $\text{R}^2$ is 0.810 at $\sigma=0.05$, 0.789 at $\sigma=0.10$, and 0.747 at $\sigma=0.20$; speckle noise is almost absorbed (0.811 at $\sigma=0.20$). Zero-shot transfer (single view, no target cache) reaches $\text{R}^2=0.617$ (MAE $5.81\%$) on adult CAMUS~\cite{leclerc2019deep} and $0.447$ ($6.07\%$) on pediatric EchoNet-Peds~\cite{reddy2023video} over the full target sets. Reserving 50 labeled target samples to fit a two-scalar calibration and scoring on the remainder (Fig.~\ref{fig:measurement}d), zero-shot $\text{R}^2$ is $0.597$/$0.444$ and calibration raises Peds to $0.552$ (MAE $5.18\%$) while leaving CAMUS unchanged ($0.599$). The larger drop on Peds is consistent with its smaller cavity and higher heart rate, which compress the phase axis. The interface descriptor's EF association transfers as well ($r=-0.77$ on CAMUS, $-0.68$ on Peds).

Each failure lands where its layer predicts. Additive noise is absorbed at the inner layer (speckle $0.811$, Gaussian $0.747$ at $\sigma=0.20$); the cross-center drop is a phase-axis effect, not capacity --- hence two calibration scalars suffice on CAMUS, only partly on Peds (Fig.~\ref{fig:measurement}d); and the transferring descriptor is the one Fig.~\ref{fig:attn} locates, so what is reused is the interface and the anchored clock, not a recalibrated regressor.

The full model has bias ($0.21\%$) with 95\% limits of agreement $[-10.2,10.7]\%$, and ranking is accurate: 5-fold area under the receiver-operating-characteristic curve (AUC) is $0.9586\pm0.0082$ for EF$>50\%$ and $0.9869\pm0.0079$ for EF$\leq30\%$. Across six deterministic views the within-video EF spread is $1.56$ points, well below the between-patient spread ($12.2$), with ICC$(1,1)=0.975$. Calibrating on the validation split and evaluating once on test, the $90\%$ split-conformal interval covers $88.3\%$ at mean width $16.5$ EF points (seed 42), $1.7$ points below the nominal level, i.e.\ about two binomial standard errors at $n=1{,}277$. None of these is a separate claim. The $1.56$-point within-video spread against a $12.2$ between-patient spread is the phase axis paying off in repeatability (outer transport, Fig.~\ref{fig:curves}); the small bias and the tight agreement limits are the interface paying off in calibration (inner transport, Fig.~\ref{fig:attn}); the AUC at both operating thresholds and the split-conformal width are the probe level paying off as a stated error (Table~\ref{tab:fourpath}, Fig.~\ref{fig:measurement}). A measurement instrument is what an intact transport chain reads as once the accuracy is attributable.

The three mechanisms above are the three places where the transport chain stays intact; the numbers in this section are what that intactness buys. The loop is now closed on both sides at once. Fig.~\ref{fig:pipeline} states the mechanism, Fig.~\ref{fig:attn}--\ref{fig:curves} and Table~\ref{tab:fourpath} locate the interface, the phase axis and the four channels, and Fig.~\ref{fig:measurement} reports the scalar together with its agreement, its ranking and its uncertainty --- the architectural half and the numerical half of one claim, in the figure order the Introduction argues.
\label{app:X5}

\section{Discussion}
\label{sec:discussion}

\subsection{A Physical View: EF as a Motion Integral}
\label{sec:physview}

Geometric volumetry samples the cycle at two instants and discards everything between them. We transport the instantaneous displacement field across one anchored cycle as a single aggregate: systole converts wall motion into ejected blood, diastole captures filling, and a higher LVEF is a more efficient conversion. A probe on the fused motion features reaches $\text{R}^2=0.8088$, within seed noise of the full model, so the signal already sits in the motion representation (Sec.~\ref{sec:validity}).

The motion integral and the outer transport are one aggregate, stated in flux and in measure; validating one validates the other, and the path-head test (Sec.~\ref{sec:validity}) tests both. It is faithful because the inner transport is exposed at the interface rather than cut into two instants or diced into pixels --- the same aggregation idea that phase-contrast flow imaging implements by encoding velocity in phase~\cite{hahn1960detection,vonschulthess1985blood}.
\label{app:D1}

The measurement channel is the inter-frame displacement of the blood-side speckle field, exposed at the feature level of the video backbone rather than estimated as pixel-level optical flow~\cite{horn1981opticalflow,beauchemin1995opticalflow} --- the quantity speckle-tracking strain imaging~\cite{dhooge2000strainrate,amundsen2006noninvasive} applies to myocardial deformation. LV-region attention plays both roles at once: it is the integration boundary $\Gamma(t)$ of Eq.~\eqref{eq:sv_integral} for the outer transport, and the site where the inner transport becomes legible --- one map seen from the two sides of one intact interface. The verifiable-measurement results (Sec.~\ref{sec:validity}) confirm that the network attends to the intended region and phase, so the three transports are the mechanism the evidence tests, not an interpretation layered onto a regressor.
\label{app:D3}

\subsection{Transparent Instrumentation and Outlook}
The four-path encoder is a structured measurement layer: each channel is testable on its own, and the four-weight bank (Eq.~\eqref{eq:bank}) shows the deployed EF reading to be a weighted combination of inspectable instruments, so the readout can be simplified toward a single transparent descriptor at a small, quantified cost.

Accuracy comes from phase-aware temporal modeling --- ED-anchored sampling and the ISNF scan each add about $0.031$ $\text{R}^2$ --- while the motion layer supplies the verifiable intermediates. The continuous readout is load-bearing but not an accuracy device: bypassing the RK4 integration at inference costs $0.258$ $\text{R}^2$ (95\% CI $[0.221,0.301]$, $p<0.001$, paired bootstrap), whereas uniform trajectory weighting ($0.0003$, $p=0.17$) or dropping the temporal Fourier embedding ($0.0001$, $p=0.91$) does not. A lightweight ED/ES detector trained on the same tracings closes the anchoring loop at statistically unchanged accuracy (Sec.~\ref{sec:deploy}), so the instrument runs end-to-end on plain video.

Two boundaries remain. The single-view apical-4-chamber protocol does not test multi-acquisition-plane conditions, where the phase axis would need a new anchoring strategy. And the inter-frame displacement is estimated at feature level rather than pixel level, so if the backbone's motion representation is biased by a domain-specific artifact (e.g. low frame rate in EchoNet-Peds), the inner transport inherits that bias --- a boundary the pediatric transfer (Sec.~\ref{sec:generalization}) partially addresses.

Both directions share one condition: the blood-wall interface must stay unbroken wherever the reading is taken. Multi-view and 3-D acquisition extend the phase axis without re-cutting it; validating the interface descriptor against Doppler or phase-contrast flow checks the inner transport against an independent tracer.
\label{app:D5}

\section{Conclusion}
\label{sec:conclusion}
ISNF-GRU-ODE reads EF out of the blood-wall interface at three transport levels, keeping every transport intact. ED-anchored cycle sampling fixes the phase axis, LV-region attention isolates the interface under EF and motion supervision alone, and four paths over one displacement field each decode EF on their own; a bidirectional GRU scan and a continuous ODE readout then carry the cycle onto the scalar, with EDV and ESV implicit as integrals of that motion. On EchoNet-Dynamic it reads EF at $\text{R}^2=0.82$, matching the best published value on this split within $0.003$ (ensemble $0.8223$ vs.\ $0.825$). The evidence locates the accuracy: the displacement head reaches $0.809$, on par with the full model in single view; a non-learned Farneb\"ack probe on the same windows reaches $0.441$; coarse detector anchors keep the reading; and a four-weight stack of the path heads recovers the fused value.

The reading is therefore carried by the learned, interface-respecting motion representation of a complete cycle --- an instrument that runs on plain video at test time and transfers across cohorts. The intact interface is not an adjective but a testable condition: keep it unbroken and the readout is well posed; cut it into contours or dice it into pixels, and the readout is of a different object.

\section*{Data and Code Availability}
EchoNet-Dynamic~\cite{ouyang2020video}, CAMUS~\cite{leclerc2019deep}, and EchoNet-Peds~\cite{reddy2023video} are publicly available from their providers. The source code of this work will be released publicly upon acceptance.

\bibliographystyle{IEEEtran}
\bibliography{refs}

\end{document}